\documentclass[showpacs,aps,prc,twocolumn,superscriptaddress]{revtex4}

\usepackage{graphicx}
\usepackage{dcolumn}
\usepackage{bm}

\begin{document}


\title{Coulomb excitation of $^{68}_{28}$Ni$_{40}$ at safe energies}

\author{N. Bree}
\affiliation{Instituut voor Kern- en Stralingsfysica, K.U.Leuven,
Celestijnenlaan 200D, B-3001 Leuven, Belgium}
\author{I. Stefanescu}
\affiliation{Instituut voor Kern- en Stralingsfysica, K.U.Leuven,
Celestijnenlaan 200D, B-3001 Leuven, Belgium}
\author{P. A. Butler}
\affiliation{Oliver Lodge Laboratory, University of Liverpool,
United Kingdom}
\author{J. Cederk\"all}
\affiliation{ISOLDE, CERN, CH-1211 Geneva 23, Switzerland}
\author{T. Davinson}
\affiliation{Department of Physics and Astronomy, University of
Edinburgh, United Kingdom}
\author{P. Delahaye}
\affiliation{ISOLDE, CERN, CH-1211 Geneva 23, Switzerland}
\author{J. Eberth}
\affiliation{IKP, University of Cologne, D-50937 Cologne, Germany}
\author{D. Fedorov}
\affiliation{Petersburg Nuclear Physics Institute, 188300
Gatchina, Russia}
\author{V. N. Fedosseev}
\affiliation{ISOLDE, CERN, CH-1211 Geneva 23, Switzerland}
\author{L. M. Fraile}
\affiliation{ISOLDE, CERN, CH-1211 Geneva 23, Switzerland}
\author{S. Franchoo}
\affiliation{IPN Orsay, F-91406 Orsay Cedex, France}
\author{G. Georgiev}
\affiliation{ISOLDE, CERN, CH-1211 Geneva 23, Switzerland}
\affiliation{CSNSM, CNRS/IN2P3; Universit\'e Paris-Sud, UMR8609,
F-91405 ORSAY-Campus, France}
\author{K. Gladnishki}
\affiliation{Dipartimento di Fisica, Universit\`a di Camerino,
I-62032 Camerino, Italy}
\author{M. Huyse}
\affiliation{Instituut voor Kern- en Stralingsfysica, K.U.Leuven,
Celestijnenlaan 200D, B-3001 Leuven, Belgium}
\author{O. Ivanov}
\affiliation{Instituut voor Kern- en Stralingsfysica, K.U.Leuven,
Celestijnenlaan 200D, B-3001 Leuven, Belgium}
\author{J. Iwanicki}
\affiliation{Heavy Ion Laboratory, Warsaw University,
Pasteura 5A, 02-093 Warsaw, Poland}
\author{J. Jolie}
\affiliation{IKP, University of Cologne, D-50937 Cologne, Germany}
\author{U. K\"oster}
\affiliation{ISOLDE, CERN, CH-1211 Geneva 23, Switzerland}
\affiliation{Institut Laue Langevin, 6 rue Jules Horowitz, F-38042
Grenoble Cedex 9, France}
\author{Th. Kr\"oll}
\affiliation{Technische Universit\"at M\"unchen,
James-Franck-Str., D-85748 Garching, Germany}
\author{R. Kr\"ucken}
\affiliation{Technische Universit\"at M\"unchen,
James-Franck-Str., D-85748 Garching, Germany}
\author{B. A. Marsh}
\affiliation{ISOLDE, CERN, CH-1211 Geneva 23, Switzerland}
\author{O. Niedermaier}
\affiliation{Max-Planck-Institut f\"ur Kernphysik, Heidelberg,
Germany}
\author{P. Reiter}
\affiliation{IKP, University of Cologne, D-50937 Cologne, Germany}
\author{H. Scheit}
\affiliation{Max-Planck-Institut f\"ur Kernphysik, Heidelberg,
Germany}
\author{D. Schwalm}
\affiliation{Max-Planck-Institut f\"ur Kernphysik, Heidelberg,
Germany}
\author{T. Sieber}
\affiliation{Ludwig Maximilians Universit\"at M\"unchen, Am
Coulombwall 1, D-85748 Garching, Germany}
\author{J. Van de Walle}
\affiliation{Instituut voor Kern- en Stralingsfysica, K.U.Leuven,
Celestijnenlaan 200D, B-3001 Leuven, Belgium}
\author{P. Van Duppen}
\affiliation{Instituut voor Kern- en Stralingsfysica, K.U.Leuven,
Celestijnenlaan 200D, B-3001 Leuven, Belgium}
\author{N. Warr}
\affiliation{IKP, University of Cologne, D-50937 Cologne, Germany}
\author{D. Weisshaar}
\affiliation{IKP, University of Cologne, D-50937 Cologne, Germany}
\author{F. Wenander}
\affiliation{ISOLDE, CERN, CH-1211 Geneva 23, Switzerland}
\author{S. Zemlyanoy}
\affiliation{Joint Institute for Nuclear Research, 141980, Dubna,
Moscow Region, Russia}


\begin{abstract}

The $B(E2;0^+\rightarrow2^+)$ value in $^{68}$Ni has been measured
using Coulomb excitation at safe energies. The $^{68}$Ni
radioactive beam was postaccelerated at the CERN on-line isotope mass separator (ISOLDE) facility to 2.9 MeV/u and directed to a $^{108}$Pd target. The emitted $\gamma$ rays were detected by
the MINIBALL detector array. Not only directly registered but also indirectly deduced information on the nucleus emitting the $\gamma$ ray was used to perform the Doppler correction, leading to a larger center-of-mass angular range
to infer the excitation cross section. The obtained value of 2.8$^{+1.2}_{-1.0}\times$10$^2$ $e^2$fm$^4$ is in good agreement with the value measured at
intermediate energy Coulomb excitation, confirming the low $0^+\rightarrow2^+$
transition probability.

\end{abstract}

\maketitle

The structure of $^{68}$Ni with a closed proton shell at $Z=28$
and a subshell closure at $N=40$ has been investigated in
$\beta$ decay \cite{Mue99}, deep inelastic reactions \cite{Bro95},
Coulomb excitation \cite{Sor02} and mass measurements
\cite{Gue07}. The latter shows no evidence for a large energy gap
at $N=40$ that separates the unique parity $\nu g_{9/2}$ orbital
from the $\nu 2p_{\frac{3}{2}}1f_{\frac{5}{2}}2p_{\frac{1}{2}}$
orbitals. More recent mass measurements do show a local weak
discontinuity in the two-neutron separation energy, thus
confirming the very weak $N=40$ subshell gap \cite{Rah07}. Still,
the $^{68}$Ni nucleus possesses nuclear properties that are
characteristic for a doubly magic nucleus: a high $2^+$ energy and
a low $B(E2;0^+\rightarrow2^+)$ value
\cite{Ber82,Ish00,Roo04,Per06,Kan06}. In recent work it has been
advocated that the high $2^+$ energy is largely due to the
opposite parity of the $\nu
2p_{\frac{3}{2}}1f_{\frac{5}{2}}2p_{\frac{1}{2}}$ orbitals and the
$\nu g_{9/2}$ orbital and that a major part of the $B(E2)$
strength resides at high energy \cite{Gra01,Lan03}. Although the high
energy of the first $2^+$ state at 2033 keV has been measured by
different experiments \cite{Mue99,Bro95,Sor02}, the low
$B(E2;0^+\rightarrow2^+)$ value has been obtained from one Coulomb
excitation experiment performed in inverse kinematics and using a
$^{68}$Ni beam at intermediate energy (produced from the
fragmentation of a $^{70}$Zn beam with an energy of 65.9 MeV/u).
The value obtained was 255$\pm$60 $e^2$fm$^4$ \cite{Sor02},
approximately 2 times lower than the $B(E2;0^+\rightarrow2^+)$ in
$^{56}$Ni, with $Z=N=28$. As this low $B(E2;0^+\rightarrow2^+)$
value is crucial for understanding the structure of $^{68}$Ni, a
new experiment aimed at measuring this value was performed using a
postaccelerated $^{68}$Ni beam from the CERN on-line isotope mass separator (ISOLDE) facility.

In this note, we report on a determination of the
$B(E2;0^+\rightarrow2^+)$ value of $^{68}$Ni using safe Coulomb
excitation where the contribution of nuclear effects in the
excitation process is limited because the separation between the
surfaces of the colliding nuclei does not drop below 5 fm over the
detected scattering range \cite{Cli86}. The $^{68}$Ni ($T_{1/2}=$
29 s) ion beam was produced at the ISOLDE radioactive-beam
facility by bombarding a 1.4-GeV proton beam, produced by the PS
booster accelerator, on a UC$_x$ target of 52 g/cm$^2$. After
diffusion of the fission products from the target and transport to
the ion source, the nickel atoms were selectively laser ionized
\cite{Mis93,Koe03,Fed00} and mass separated, yielding an
average beam intensity of approximately $2.5\times 10^{6}$
particles per second at 60 keV \cite{Koe}. Subsequently, the beam
was postaccelerated by REX-ISOLDE \cite{Hab98} up to an energy of
2.9 MeV/u. Coulomb excitation was induced by directing the
postaccelerated $^{68}$Ni beam at $v/c \sim 0.08$ to a 2
mg/cm$^2$ $^{108}$Pd target. The scattered nuclei were detected by
a double sided silicon strip detector (DSSSD) \cite{Ost02},
consisting of four quadrants, mounted at a distance of 32 mm
behind the target position. Each quadrant was divided in 16
annular strips at the front side and 24 sector strips at the back
side. The detection range in the laboratory frame covered angles
from 16$^{\circ}$ up to 53$^{\circ}$. Eight MINIBALL \cite{Ebe01}
clusters, consisting of three sixfold segmented HPGe crystals,
were placed around the target chamber at a distance of
approximately 12 cm from the target position to detect the emitted
$\gamma$ rays. An add-back procedure was applied, summing up the
$\gamma$ rays simultaneously incident on crystals of the same
cluster and therefore correcting for Compton scattering within
the same cluster. At an energy of 1332 keV the absolute full
energy peak efficiency after applying the add-back procedure was
7.1$\%$ and at 2033 keV it was 5.6$\%$. The total measuring time was
41 hours, the beam intensity of $^{68}$Ni on the $^{108}$Pd target
throughout the experiment being $\sim10^{4}$ particles per second
on average, using the production sequence as explained below.
Similar Coulomb excitation experiments with this setup at
comparable energies and masses have been successfully performed for the study
of neutron-rich Cu and Zn isotopes \cite{Ste07,Van07}.

The only isobaric contaminant present in the beam was $^{68}$Ga
($T_{1/2}=$ 67.63 m), which, due to its relatively low ionization
potential, was ionized at the high-temperature surface of the ion
source. Moreover, the release time of $^{68}$Ga from the target
and ion source system (in the range of a few seconds) is much
shorter than that of $^{68}$Ni (in the range of a few minutes)
\cite{Koe}. To reduce the overwhelming contamination,
considerations about the time structure of the beam are of crucial
importance. The proton beam from the PS booster comes in
supercycles consisting of 12 proton pulses per cycle. The
individual pulses ($3\times 10^{13}$ protons per pulse) are
separated by a time interval of 1.2 s. Having protons on the
UC$_x$ primary target while acquiring data would cause the beam to
consist mostly of the dominant $^{68}$Ga contaminant, and
$^{68}$Ni would be a rather small component of the beam incident
on the secondary $^{108}$Pd target.
\begin{figure}[h!]
\includegraphics[scale=0.37]{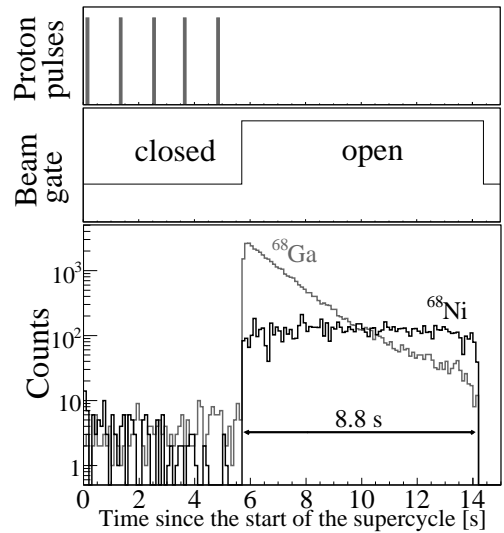}
\caption{\label{fig:sstd} Release curves for $^{68}$Ni and the
$^{68}$Ga isobaric contaminant as a function of the time difference with the start of the supercycle. 800 ms after the fifth proton pulse the beam gate is opened, and the produced ions are sent to the experimental setup.}
\end{figure}

Due to the strong difference in release time and the pulsed proton
beam structure, a specific technique can be used. For the present
experiment the first five pulses of the supercycle were taken to
produce the radioactive nuclei, and the ion beam gate was kept
closed up to this moment (see Fig. \ref{fig:sstd}). Only 800 ms
after the fifth pulse the beam gate was opened, causing a
considerable reduction of the rapidly releasing $^{68}$Ga, whereas
the $^{68}$Ni isotopes were still diffusing out of the primary target ion
source system with a constant rate. A similar technique has been
applied in the heavy lead region \cite{Dup98}. To determine the beam
composition, scattering data in the DSSSD detector were compared
between periods when the laser beams were on ($^{68}$Ga and
$^{68}$Ni) and off ($^{68}$Ga only). A shutter was periodically
shifted in front of the laser beam to prevent the laser light from
reaching the ion source during every second supercycle. Hence,
``laser on" and ``laser off" data have been subsequently acquired.
Figure \ref{fig:sstd} shows the number of elastically scattered
particles in the DSSSD detector as a function of the time
difference with the start of the 14.4 seconds long supercycle. The
$^{68}$Ga information has been obtained using the laser off data,
whereas subtracting the laser off from the laser on data results in
a pure $^{68}$Ni curve. Although the $^{68}$Ga contaminant is
released following a quasiexponential function, $^{68}$Ni is
coming out almost constantly, due to the long release time of the
element. The timing sequence of the proton pulses and the status
of the beam gate are shown in Fig. \ref{fig:sstd} as well. This setting resulted in a
$^{68}$Ni to total beam intensity ratio of 24$\pm$1$\%$. The
actual measuring time when the $^{68}$Ni beam was impinging on
the target amounted to 8.8 seconds per supercycle (14.4-5.6
seconds). When only selecting the last 5 s of the
supercycle, the $^{68}$Ga contamination is lowered, enhancing the
ratio of the $^{68}$Ni beam component to the full beam intensity
to 68$\pm$1$\%$.
\begin{figure*}
\includegraphics[scale=0.42]{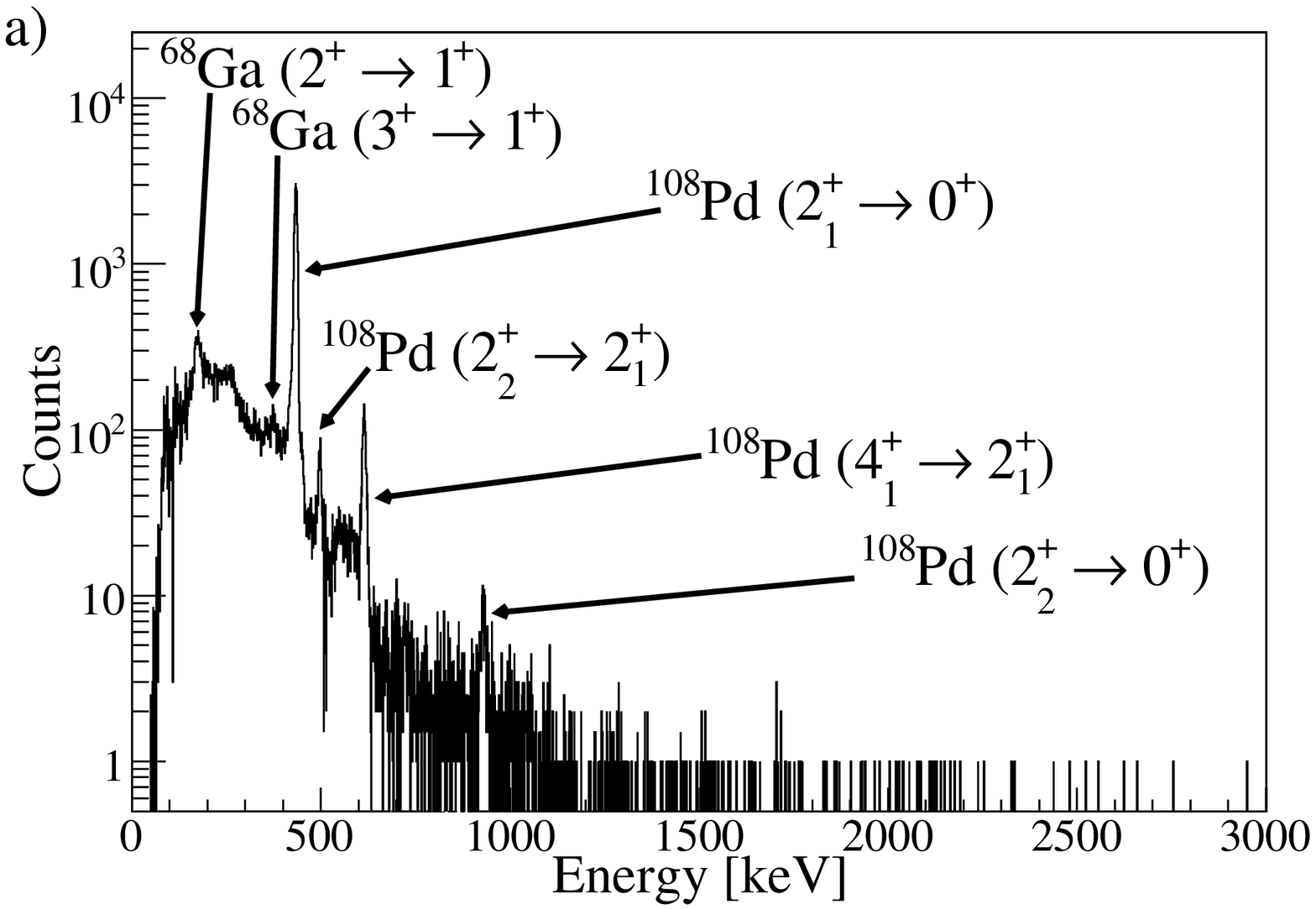}
\includegraphics[scale=0.42]{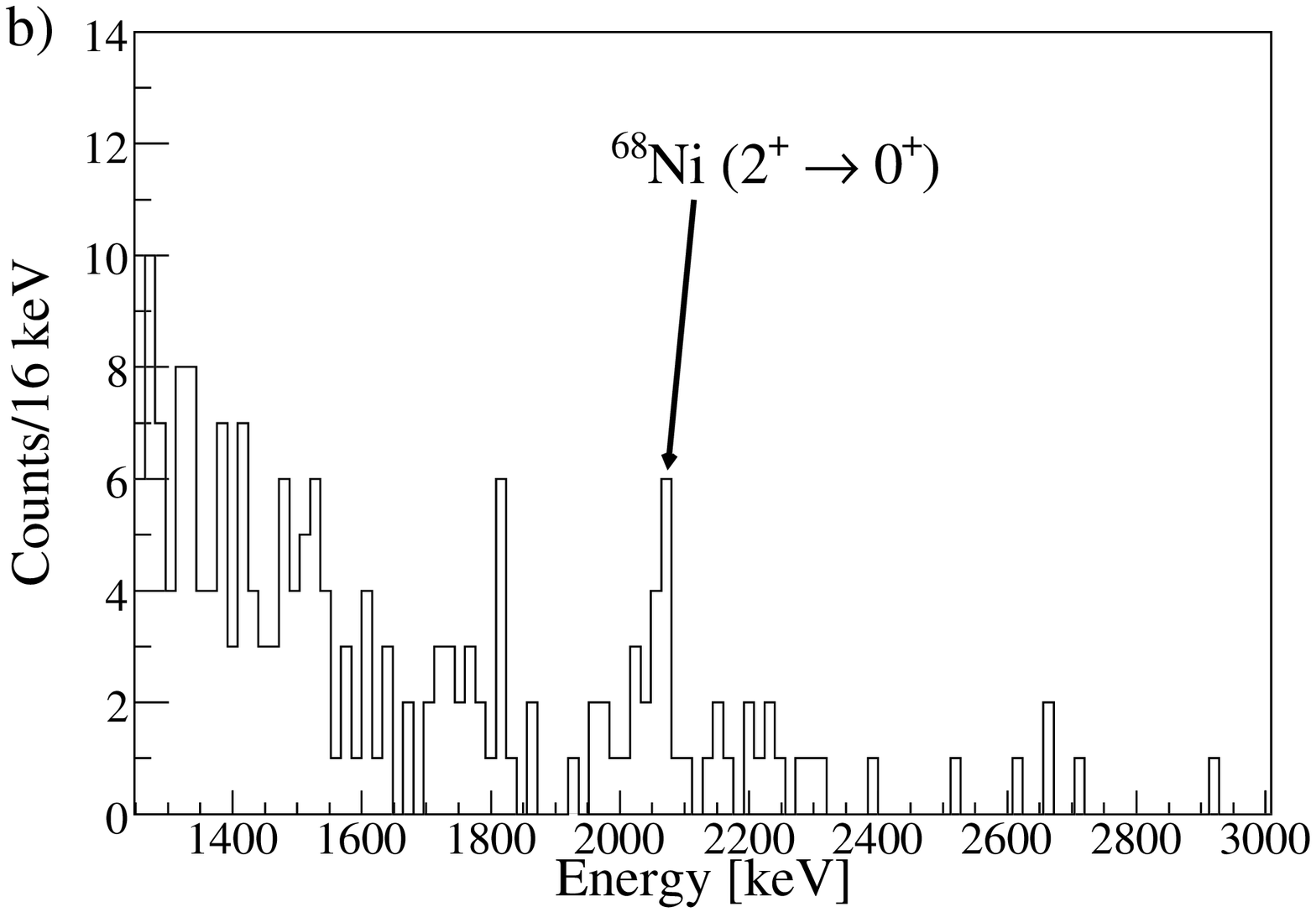}
\caption{\label{fig:all} $\gamma$-ray spectrum with particle
coincidence Doppler corrected for $^{108}$Pd target excitation (a)
and part of the $\gamma$-ray spectrum with particle coincidence
Doppler corrected for $^{68}$Ni projectile excitation (b) focused
around 2033 keV.}
\end{figure*}

\begin{figure}[h!]
\includegraphics[scale=0.41]{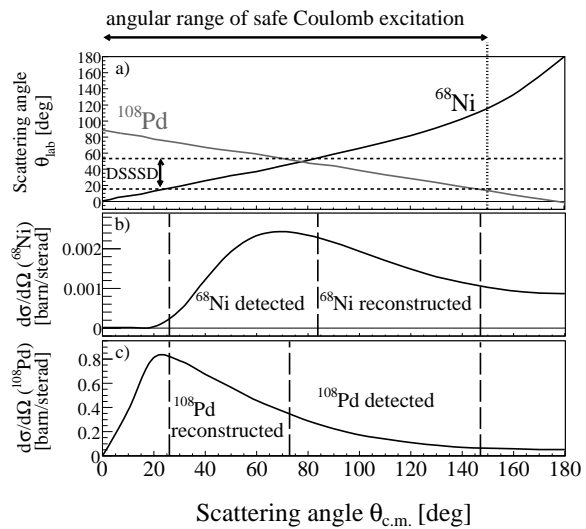}
\caption{\label{fig:angles3} Scattering angles of the $^{68}$Ni
projectile and the $^{108}$Pd target particles in the laboratory
frame versus the scattering angle in the center-of-mass frame (a).
From the angular range of the DSSSD detector it can be seen where the information on the $\gamma$-emitting nucleus was registered directly and where it was deduced indirectly.
Also the differential cross sections for both nuclei are shown
(b and c). All detected events occur in the angular range of safe
Coulomb excitation, where the distance between the nuclear
surfaces never exceeds 5 fm.}
\end{figure}

Figure \ref{fig:all} shows the $\gamma$-ray spectra coincident
with a particle detected in the DSSSD and Doppler corrected for
detection of the $^{108}$Pd target recoil (a) and the $^{68}$Ni
scattered projectile (b), respectively, with data taken during the 8.8
s of the supercycle. The latter was zoomed around the energy
of the first excited 2$^+$ state at 2033 keV. The Doppler-corrected spectrum for $^{108}$Pd clearly shows the $\gamma$ lines
from the population of the 2$^+_1$, 2$^+_2$, and 4$^+_1$ states in
the target. Two smeared out photo peaks from the transitions in
the $^{68}$Ga contaminant are also visible. The $\gamma$ events
shown in both spectra are in prompt coincidence with the detection
of a particle in the silicon strip detector, which can be the
projectile or the target or both. The energy deposited in the
DSSSD detector as a function of the laboratory angle allows the
projectile ($^{68}$Ni and $^{68}$Ga) and target ($^{108}$Pd) ions
to be discriminated. The following procedure was used to perform the Doppler correction (see Fig. \ref{fig:angles3}). In case a Doppler-shifted 2033-keV $\gamma$ ray is emitted by the $^{68}$Ni projectile, knowledge on the direction and the energy of the scattered ion is required to perform the Doppler correction. If the $^{68}$Ni nucleus is detected in the DSSSD detector, these quantities are registered in a direct way.
This occurs when the projectile has been scattered between 16$^{\circ}$
and 53$^{\circ}$ in the laboratory system (26$^{\circ}$ and
83$^{\circ}$ in the center-of-mass frame). In case a $^{108}$Pd
target particle was detected in the silicon strip detector, the
$^{68}$Ni projectile was scattered between 46 $^{\circ}$ and
112$^{\circ}$ in the laboratory system (73$^{\circ}$ and
148$^{\circ}$ in the center-of-mass frame). Only in the small
range between 46 $^{\circ}$ and 53$^{\circ}$ (73$^{\circ}$ and
83$^{\circ}$ in the center-of-mass frame) the projectile was
detected as well, and this happens of course in opposite
quadrants. If the angle of the projectile is higher, the registered position and energy of the associated target recoil of the collision were used to infer the direction and the energy of the scattered $^{68}$Ni projectile, assuming inelastic scattering in two-body kinematics, hereby correcting for the energy loss in the target. Applying this method enlarges substantially the center-of-mass angular range used to integrate the cross section, increasing the
statistics in the particle-coincident $\gamma$-ray spectrum with a
factor of 2.7 for $^{68}$Ni and 2.6 for $^{108}$Pd. Figure
\ref{fig:angles3} shows how the scattering angles of projectile and target are correlated in the
laboratory frame as a function of the
angle in the center-of-mass frame, indicating the angular range
covered by the DSSSD detector. Parts of the angular range
outside the detection range of the DSSSD are now included by indirectly deriving direction and energy of the nondetected nucleus emitting the $\gamma$ ray. All events registered occur at a center-of-mass angle lower than 150$^{\circ}$, insuring the safe nature
of the Coulomb excitation process. When exceeding this angle,
nuclear effects can no longer be excluded, as the distance
between the surfaces of the two colliding nuclei becomes smaller
than 5 fm.

In Fig. \ref{fig:all}(b) a peak structure of 11 counts is visible
around 2 MeV on a constant background of 1 count per 16 keV energy bin. The
photo peak energy 2050$\pm$65 keV corresponds well to the known
value of 2033 keV. This peak is therefore assigned to the
$2^+\rightarrow0^+$ transition in $^{68}$Ni resulting in a total
number of counts 11$^{+5}_{-4}$. The asymmetry in the
error bar is due to the low number of counts,
implying the use of Poisson statistics \cite{Rol01}.
As the figure shows only the high-energy events, the strongest Coulomb excitation lines of $^{68}$Ga,
which are for this odd-odd nucleus mainly of low-energy nature,
are not visible. A separate article will discuss the Coulomb
excitation information on the Ga contaminant.

\begin{figure}
\includegraphics[scale=0.39]{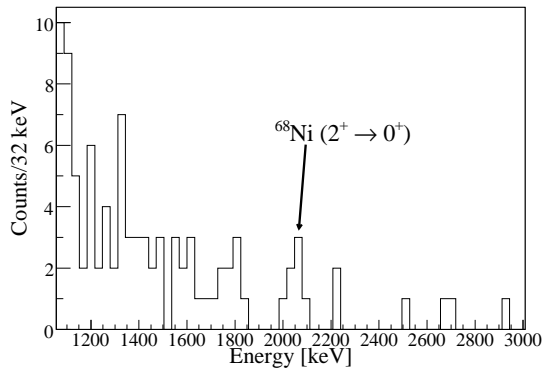}
\caption{\label{fig:104sstd} Particle-coincident $\gamma$-ray
spectrum Doppler corrected for $^{68}$Ni. Only the events occurring within the last 5 s of the proton pulse supercycle are included.}
\end{figure}

A consistency check was performed by selecting the
$\gamma$-ray events occurring within the last 5 s of the
supercycle, resulting in a suppression of the $^{68}$Ga lines, and
therefore decreasing the background. Figure \ref{fig:104sstd}
shows the $\gamma$-ray events with the same conditions as the
$^{68}$Ni spectrum in Fig. \ref{fig:all}, but with this time
selection in the supercycle. Integrating the photo peak at 2033
keV in this $^{68}$Ni-enhanced spectrum results in 7 prompt counts
and 1 count in the background, yielding a rate of 6$^{+3}_{-2}$ counts.
The time selection (of the last 5 s) covers 57$\%$ of the
total time of the supercycle when the beam was incident on the
$^{108}$Pd target (8.8 s). Due to the fact that $^{68}$Ni is
released constantly, the amount of excitations in the nucleus
should decrease by approximately 57$\%$ when using the time
selection. The counting rate of 6$^{+3}_{-2}$ in the $^{68}$Ni-enhanced
spectrum is therefore consistent with the 11$^{+5}_{-4}$ de-excitations
seen in the $\gamma$-ray spectrum without a time selection in the
supercycle. The amount of $^{68}$Ga within this supercycle time
selection is 8$\%$ compared to the full statistics, as can be
calculated from the elastically scattered particles in the DSSSD
detector during laser on/off data (Fig. \ref{fig:sstd}).
Comparing the $\gamma$-ray intensity of $2^{+} \rightarrow 1^{+}$
transition in $^{68}$Ga results in a value of 7$\%$, again
confirming the consistency of this method.

Of the 11$^{+5}_{-4}$ counts in the de-excitation photo peak, the
cross section for exciting $^{68}$Ni to its first 2$^+$ state was
determined using the program \textit{GOSIA2} \cite{Czo05}, which
calculates the $\gamma$-ray yields after integration over the
scattering ranges and energy loss in the target, taking into
account the angular distribution of the $\gamma$ rays. Hereby the quadrupole moment of the 2$^+$ state was assumed to be small and the reorientation effect was neglected. The
amount of de-excitations in the photo peak of the known
$2^+\rightarrow0^+$ transition in $^{108}$Pd \cite{Sve95} was used
for normalization, correcting it for $^{68}$Ga-induced excitation.
The uncertainty on the count rate in the target de-excitation photo
peak contributed only 3$\%$ to the total error.
The experimentally measured $B(E2)$ value in $^{68}$Ni was
2.8$^{+1.2}_{-1.0}\times$10$^2$ $e^2$fm$^4$. This result is less accurate but
in good agreement with the value measured using Coulomb excitation
at intermediate energy (255$\pm$60 $e^2$fm$^4$), confirming the low transition probability
of 260$\pm$50 $e^2$fm$^4$ from a weighted average.

Studies involving safe Coulomb excitation experiments on more
neutron-rich nickel isotopes are not hindered by the in-target
production using proton-induced fission but by the slow release
out of the target matrix and the strong contamination of the
gallium isobars. Vigorous research is needed to bring down this
delay time to the required seconds range and to reduce the gallium contamination to extend this study toward the doubly
magic $^{78}$Ni nucleus. Work in this direction is underway at
ISOLDE \cite{Fed00}.

This work was supported by EU Sixth Framework through No.
EURONS-506065, BMBF under Contract No. 06KY205I and No. 06MT238,
BriX IAP Research Program No. P6/23, and FWO-Vlaanderen (Belgium). We acknowledge fruitful discussions with K. Riisager.
\bibliographystyle{apsrev}
\bibliography{Ni}

\begin{thebibliography}{26}
\expandafter\ifx\csname natexlab\endcsname\relax\def\natexlab#1{#1}\fi
\expandafter\ifx\csname bibnamefont\endcsname\relax
  \def\bibnamefont#1{#1}\fi
\expandafter\ifx\csname bibfnamefont\endcsname\relax
  \def\bibfnamefont#1{#1}\fi
\expandafter\ifx\csname citenamefont\endcsname\relax
  \def\citenamefont#1{#1}\fi
\expandafter\ifx\csname url\endcsname\relax
  \def\url#1{\texttt{#1}}\fi
\expandafter\ifx\csname urlprefix\endcsname\relax\def\urlprefix{URL }\fi
\providecommand{\bibinfo}[2]{#2}
\providecommand{\eprint}[2][]{\url{#2}}

\bibitem[{\citenamefont{Mueller et~al.}(1999)}]{Mue99}
\bibinfo{author}{\bibfnamefont{W.}~\bibnamefont{Mueller}} \bibnamefont{et~al.},
  \bibinfo{journal}{Phys. Rev. Lett.} \textbf{\bibinfo{volume}{83}},
  \bibinfo{pages}{3613} (\bibinfo{year}{1999}).

\bibitem[{\citenamefont{Broda et~al.}(1995)}]{Bro95}
\bibinfo{author}{\bibfnamefont{R.}~\bibnamefont{Broda}} \bibnamefont{et~al.},
  \bibinfo{journal}{Phys. Rev. Lett.} \textbf{\bibinfo{volume}{74}},
  \bibinfo{pages}{868} (\bibinfo{year}{1995}).

\bibitem[{\citenamefont{Sorlin et~al.}(2002)}]{Sor02}
\bibinfo{author}{\bibfnamefont{O.}~\bibnamefont{Sorlin}} \bibnamefont{et~al.},
  \bibinfo{journal}{Phys. Rev. Lett.} \textbf{\bibinfo{volume}{88}},
  \bibinfo{pages}{092501} (\bibinfo{year}{2002}).

\bibitem[{\citenamefont{Gu\'enaut et~al.}(2007)}]{Gue07}
\bibinfo{author}{\bibfnamefont{C.}~\bibnamefont{Gu\'enaut}}
  \bibnamefont{et~al.}, \bibinfo{journal}{Phys. Rev. C}
  \textbf{\bibinfo{volume}{75}}, \bibinfo{pages}{044303}
  (\bibinfo{year}{2007}).

\bibitem[{\citenamefont{Rahaman et~al.}(2007)}]{Rah07}
\bibinfo{author}{\bibfnamefont{S.}~\bibnamefont{Rahaman}} \bibnamefont{et~al.},
  \bibinfo{journal}{Eur. Phys. J. A} \textbf{\bibinfo{volume}{34}},
  \bibinfo{pages}{5} (\bibinfo{year}{2007}).

\bibitem[{\citenamefont{Bernas et~al.}(1982)}]{Ber82}
\bibinfo{author}{\bibfnamefont{M.}~\bibnamefont{Bernas}} \bibnamefont{et~al.},
  \bibinfo{journal}{Phys. Lett. B} \textbf{\bibinfo{volume}{113}},
  \bibinfo{pages}{279} (\bibinfo{year}{1982}).

\bibitem[{\citenamefont{Ishii et~al.}(2000)}]{Ish00}
\bibinfo{author}{\bibfnamefont{T.}~\bibnamefont{Ishii}} \bibnamefont{et~al.},
  \bibinfo{journal}{Phys. Rev. Lett.} \textbf{\bibinfo{volume}{84}},
  \bibinfo{pages}{39} (\bibinfo{year}{2000}).

\bibitem[{\citenamefont{Van~Roosbroeck et~al.}(2004)}]{Roo04}
\bibinfo{author}{\bibfnamefont{J.}~\bibnamefont{Van~Roosbroeck}}
  \bibnamefont{et~al.}, \bibinfo{journal}{Phys. Rev. C}
  \textbf{\bibinfo{volume}{69}}, \bibinfo{pages}{034313}
  (\bibinfo{year}{2004}).

\bibitem[{\citenamefont{Perru et~al.}(2006)}]{Per06}
\bibinfo{author}{\bibfnamefont{O.}~\bibnamefont{Perru}} \bibnamefont{et~al.},
  \bibinfo{journal}{Phys. Rev. Lett.} \textbf{\bibinfo{volume}{96}},
  \bibinfo{pages}{232501} (\bibinfo{year}{2006}).

\bibitem[{\citenamefont{Kaneko et~al.}(2006)}]{Kan06}
\bibinfo{author}{\bibfnamefont{K.}~\bibnamefont{Kaneko}} \bibnamefont{et~al.},
  \bibinfo{journal}{Phys. Rev. C} \textbf{\bibinfo{volume}{74}},
  \bibinfo{pages}{024321} (\bibinfo{year}{2006}).

\bibitem[{\citenamefont{Grawe and Lewitowicz}(2001)}]{Gra01}
\bibinfo{author}{\bibfnamefont{H.}~\bibnamefont{Grawe}} \bibnamefont{and}
  \bibinfo{author}{\bibfnamefont{M.}~\bibnamefont{Lewitowicz}},
  \bibinfo{journal}{Nucl. Phys. A} \textbf{\bibinfo{volume}{693}},
  \bibinfo{pages}{116} (\bibinfo{year}{2001}).

\bibitem[{\citenamefont{Langanke et~al.}(2003)}]{Lan03}
\bibinfo{author}{\bibfnamefont{K.}~\bibnamefont{Langanke}}
  \bibnamefont{et~al.}, \bibinfo{journal}{Phys. Rev. C}
  \textbf{\bibinfo{volume}{67}}, \bibinfo{pages}{044314}
  (\bibinfo{year}{2003}).

\bibitem[{\citenamefont{Cline}(1986)}]{Cli86}
\bibinfo{author}{\bibfnamefont{D.}~\bibnamefont{Cline}}, \bibinfo{journal}{Ann.
  Rev. Nucl. Part. Sci.} \textbf{\bibinfo{volume}{36}}, \bibinfo{pages}{683}
  (\bibinfo{year}{1986}).

\bibitem[{\citenamefont{Mishin et~al.}(1993)}]{Mis93}
\bibinfo{author}{\bibfnamefont{V.}~\bibnamefont{Mishin}} \bibnamefont{et~al.},
  \bibinfo{journal}{Nucl. Instr. and Meth. B} \textbf{\bibinfo{volume}{73}},
  \bibinfo{pages}{550} (\bibinfo{year}{1993}).

\bibitem[{\citenamefont{K{\"o}ster et~al.}(2003)}]{Koe03}
\bibinfo{author}{\bibfnamefont{U.}~\bibnamefont{K{\"o}ster}}
  \bibnamefont{et~al.}, \bibinfo{journal}{Spectrochimica Acta B.}
  \textbf{\bibinfo{volume}{58}}, \bibinfo{pages}{1047} (\bibinfo{year}{2003}).

\bibitem[{\citenamefont{Fedoseyev et~al.}(2000)}]{Fed00}
\bibinfo{author}{\bibfnamefont{V.}~\bibnamefont{Fedoseyev}}
  \bibnamefont{et~al.}, \bibinfo{journal}{Hyp. Inter.}
  \textbf{\bibinfo{volume}{127}}, \bibinfo{pages}{109} (\bibinfo{year}{2000}).

\bibitem[{\citenamefont{K{\"o}ster}(2000)}]{Koe}
\bibinfo{author}{\bibfnamefont{U.}~\bibnamefont{K{\"o}ster}}, Ph.D. thesis,
  \bibinfo{school}{T. U. M{\"u}nchen} (\bibinfo{year}{2000}).

\bibitem[{\citenamefont{Habs et~al.}(1998)}]{Hab98}
\bibinfo{author}{\bibfnamefont{D.}~\bibnamefont{Habs}} \bibnamefont{et~al.},
  \bibinfo{journal}{Nucl. Instr. and Meth. B} \textbf{\bibinfo{volume}{139}},
  \bibinfo{pages}{128} (\bibinfo{year}{1998}).

\bibitem[{\citenamefont{Ostrowski et~al.}(2002)}]{Ost02}
\bibinfo{author}{\bibfnamefont{A.~N.} \bibnamefont{Ostrowski}}
  \bibnamefont{et~al.}, \bibinfo{journal}{Nucl. Instr. and Meth. A}
  \textbf{\bibinfo{volume}{480}}, \bibinfo{pages}{448} (\bibinfo{year}{2002}).

\bibitem[{\citenamefont{Eberth et~al.}(2001)}]{Ebe01}
\bibinfo{author}{\bibfnamefont{J.}~\bibnamefont{Eberth}} \bibnamefont{et~al.},
  \bibinfo{journal}{Prog. Part. Nucl. Phys.} \textbf{\bibinfo{volume}{46}},
  \bibinfo{pages}{389} (\bibinfo{year}{2001}).

\bibitem[{\citenamefont{Stefanescu et~al.}(2007)}]{Ste07}
\bibinfo{author}{\bibfnamefont{I.}~\bibnamefont{Stefanescu}}
  \bibnamefont{et~al.}, \bibinfo{journal}{Phys. Rev. Lett.}
  \textbf{\bibinfo{volume}{98}}, \bibinfo{pages}{122701}
  (\bibinfo{year}{2007}).

\bibitem[{\citenamefont{Van~de Walle et~al.}(2007)}]{Van07}
\bibinfo{author}{\bibfnamefont{J.}~\bibnamefont{Van~de Walle}}
  \bibnamefont{et~al.}, \bibinfo{journal}{Phys. Rev. Lett.}
  \textbf{\bibinfo{volume}{99}}, \bibinfo{pages}{142501}
  (\bibinfo{year}{2007}).

\bibitem[{\citenamefont{Van~Duppen et~al.}(1998)}]{Dup98}
\bibinfo{author}{\bibfnamefont{P.}~\bibnamefont{Van~Duppen}}
  \bibnamefont{et~al.}, \bibinfo{journal}{Nucl. Instr. and Meth. B}
  \textbf{\bibinfo{volume}{134}}, \bibinfo{pages}{267} (\bibinfo{year}{1998}).

\bibitem[{\citenamefont{Rolke et~al.}(2001)}]{Rol01}
\bibinfo{author}{\bibfnamefont{W.~A.} \bibnamefont{Rolke}}
  \bibnamefont{et~al.}, \bibinfo{journal}{Nucl. Instr. and Meth. A}
  \textbf{\bibinfo{volume}{458}}, \bibinfo{pages}{745} (\bibinfo{year}{2001}).

\bibitem[{\citenamefont{Czosnyka}(2005)}]{Czo05}
\bibinfo{author}{\bibfnamefont{T.}~\bibnamefont{Czosnyka}},
  \bibinfo{howpublished}{computer code GOSIA2, University of Warsaw}
  (\bibinfo{year}{2005}).

\bibitem[{\citenamefont{Svensson et~al.}(1995)}]{Sve95}
\bibinfo{author}{\bibfnamefont{L.}~\bibnamefont{Svensson}}
  \bibnamefont{et~al.}, \bibinfo{journal}{Nucl. Phys. A}
  \textbf{\bibinfo{volume}{584}}, \bibinfo{pages}{547} (\bibinfo{year}{1995}).

\end{thebibliography}

\end{document}